\documentclass[12pt,a4paper]{article}
\usepackage{amssymb}

\usepackage{float}
\usepackage{makeidx}
\usepackage[dvips]{graphics}
\usepackage[dvips]{epsfig}
\usepackage{amsmath}

\begin{document}

\title{A possible origin of superconducting currents in cosmic strings}
\author{Helder Ch\'{a}vez$^{\text{ }\ddagger \text{ }}\thanks{%
Fellow of Centro Latinoamericano de F\'{\i}sica.}$ ,{\large \ }Luis Masperi$%
^{\maltese \text{ }}$\thanks{%
On leave of absence from Centro At\'{o}mico Bariloche, San Carlos de
Bariloche, Argentina. Author to whom correspondence should be addressed.} \\
$^{\ddagger }$ Centro Brasileiro de Pesquisas F\'{\i}sicas,\ \ \\
Dr. Xavier Sigaud 150, 22290-180 Rio de Janeiro, Brazil\ \\
$^{\maltese }$ Centro Latinoamericano de F\'{\i}sica,\ \ \\
Av. Venceslau Br\'{a}z 71 Fundos, 22290-140 Rio de Janeiro, Brazil\ \\
}
\maketitle

\begin{abstract}
The scattering and capture of right-handed neutrinos by an Abelian\ cosmic
string in the$\;SO(10)$\ grand unification model are considered. The
scattering cross-section of neutrinos per unit length due to the interaction
with the gauge and Higgs fields of the string is much larger in its scaling
regime than in the friction one because of the larger infrared cutoff of the
former.The probability of capture in a zero mode of the string accompanied
by the emission of a gauge or Higgs boson shows a resonant peak for neutrino
momentum of the order of its mass. \ Considering the decrease of number of
strings per unit comoving volume in the scaling epoch the cosmological
consequences of the superconducting strings formed in this regime will be
much smaller than those which could be produced already in the friction one.
\end{abstract}

\section{Introduction}

\noindent It is possible that the early universe suffered a sequence of
phase transitions breaking symmetries of the grand unification theories
(GUT) and generating topological defects \cite{1} like the cosmic strings.

The Aharonov-Bohm effect has been described \cite{2} for the interaction of
fermions of charge $e$ with the string gauge field when the ratio $\frac{e}{%
e_{o}}$ is half-integer, being $e_{o}$ the charge of the Higgs field
responsible for the breaking of the Abelian symmetry which generates the
defect.

The corresponding cross-section diverges in the forward and backward
directions due to the long range nature of the interaction which gives
contributions even for wave packets passing at infinite distance from the
string. However the unitarity of the S-matrix has been proved \cite{3} and
the cross-section can be taken as finite introducing a cutoff for the
distance from the string.

This cutoff has the physical motivation in the correlation length between
strings. At the beginning of the so called friction regime after their
formation $\xi \simeq \frac{1}{\lambda \eta }$ where $\lambda $ is the
coupling constant of the Higgs potential which breaks the symmetry at the
energy scale $\eta .$ Afterwards, if the scaling regime is achieved $\xi
\backsim t$ so that at any time strings represent the same fraction of the
universe energy density.

The purpose of our work is twofold. On one side to include the effect of the
string Higgs field in the scattering whose order of magnitude has been
estimated \cite{4} without considering the simultaneous interaction with the
gauge field. On the other hand we wish to analyze the capture of fermions by
the string to form \ superconducting currents if there are bound zero modes
for the Dirac equation in the plane transverse to it \cite{5} . For this
last process the fermion capture has been considered \cite{4} with the
simultaneous emission of a Higgs particle, whereas we also include the
alternative emission of a gauge boson.

The simplest fermionic candidate suitable for our analysis is the
right-handed neutrino $\nu _{R}$ , which is a $SU(5)$ singlet in the
representation $\mathbf{16}$ of $SO(10)$ and that acquires mass through a
Majorana coupling with a $SU(5)$ singlet Higgs in the representation $%
\mathbf{126}$ of $SO(10)$ , responsible for the breaking of the Abelian $%
\widetilde{U}(1)$ contained in the latter. Therefore $\nu _{R}$ is the only
fermion which may form a zero mode when it is captured by the string
generated at the breaking of $\widetilde{U}(1)\ .$

In Sect. 2 we describe the Abelian string in $SO(10)$ and the relevant
fermionic field. In Sect. 3 the cross-section for scattering of Majorana
neutrinos is calculated perturbatively in the approximation of large
momentum both in the friction and scaling string regimes. Sect. 4 is devoted
to the capture of fermion with the emission either of a Higgs or a gauge
boson, finding in both cases a resonant peak for comparatively low momentum.
Sect. 5 includes conclusions and cosmological implications of a larger
influence of superconducting strings formed in the friction epoch.

\section{SO(10) Abelian cosmic strings and Majorana fermions}

The first GUT symmetry which contains an Abelian group $\widetilde{U}(1)$
additional to the electromagnetic one is $SO(10)$ \ which can be broken
according to the scheme

\begin{gather}
SO(10)\underset{\mathbf{45}}{\longrightarrow }SU(5)\otimes \widetilde{U}(1)%
\underset{\mathbf{126}}{\longrightarrow }SU(5)\otimes Z_{2}\underset{\mathbf{%
45}}{\longrightarrow }SU(3)_{C}\otimes SU(2)_{L}\otimes \widetilde{U}%
(1)_{Y}\otimes Z_{2}  \notag \\
\underset{\mathbf{10}}{\longrightarrow }SU(3)_{C}\otimes U(1)_{em}\otimes
Z_{2}\   \tag{1}
\end{gather}
where the representations of the relevant Higgs fields are indicated. The
expectation value of the Higgs field $\Phi $ in $\mathbf{126}$ breaks $%
\widetilde{U}(1)$ producing Abelian strings which are topologically stable
because the conserved discrete symmetry $Z_{2}$ avoids their breaking by
monopoles. The expectation value of $\Phi $ will be of the GUT order $\eta
\backsim 10^{15}GeV$ \ and since its $\widetilde{U}(1)$ charge is $10$
whereas that of $\nu _{R}$ denoted by $\psi $ is $5$ , a Majorana mass term
coupling which violates lepton number is possible in the Lagrangian
\begin{gather}
\mathcal{L=}\left( \mathcal{D}_{\mu }\,\Phi \right) ^{\star }\left( \mathcal{%
D}^{\mu }\,\Phi \right) -\frac{1}{4}\mathcal{F}_{\mu \nu }\mathcal{F}^{_{\mu
\nu }}-\frac{1}{4}\lambda \left( \left| \Phi \right| ^{2}-\eta ^{2}\right)
^{2}+\psi ^{\dagger }i\sigma ^{\mu }\mathcal{D}_{\mu }\psi -  \notag \\
-\frac{1}{2}\left\{ ig\psi ^{\dagger }\Phi \psi ^{c}+h.c.\right\} \ ,
\tag{2}
\end{gather}
with $\mathcal{F}_{\mu \nu }=\partial _{\mu }\mathcal{A}_{\nu }-\partial
_{\nu }\mathcal{A}_{\mu }$ , $\mathcal{D}_{\mu }\,\Phi =\left( \partial
_{\mu }-ie\mathcal{A}_{\mu }\right) \Phi $ , $\mathcal{D}_{\mu }\,\psi
=(\partial _{\mu }-\frac{1}{2}ie\mathcal{A}_{\mu })\psi ,$ $\nu
_{R}^{\,c}=i\sigma ^{2}\nu _{R}^{\star }$ , $\sigma ^{\mu }=\left( I,\sigma
^{i}\right) $ $.$ In the broken symmetry vacuum $\Phi =\eta $ and $\mathcal{A%
}_{\mu }=0$ there is a generation of masses $M_{\mathcal{H}}=\sqrt{\lambda }%
\eta $ and $M_{\mathcal{A}}=\sqrt{2}e\eta $ for the bosons and $M_{o}=g\eta $
for the fermion.

The string configuration in planar coordinates for unit winding number is
\begin{equation}
\Phi =\eta \,f\,\left( r\right) \,e^{i\varphi }\,\ ,\ \,\mathcal{A}_{\varphi
}=\frac{1}{e\,r}\,a\left( r\right)  \tag{3}
\end{equation}
with the behaviour $f\,(0)=a(0)=0$ , $f\,(\infty )=a(\infty )=1$ .

The free quantum Majorana field of right chirality \cite{6} is

\begin{gather}
\psi \left( x\right) =\frac{1}{\sqrt{\mathcal{V}}}\sum_{\overrightarrow{p}}%
\frac{1}{\sqrt{2p_{o}}}[(c\left( \mathbf{p},+\right) e^{-ip.x}+c^{\dagger
}\left( \mathbf{p},-\right) e^{\,ip.x})\sqrt{p_{o}+p}\,\chi \left(
\overrightarrow{p},+\right) +  \notag \\
(c\left( \mathbf{p},-\right) e^{-ip.x}-c^{\dagger }\left( \mathbf{p}%
,+\right) e^{\,ip.x})\sqrt{p_{o}-p}\,\chi \left( \mathbf{p},-\right) ]\ ,
\tag{4}
\end{gather}
where $\chi $ are the helicity eigenstates

\begin{equation}
\mathbf{\sigma }.\mathbf{p}\,\chi \left( \mathbf{p},\pm \right) =\pm \ p\chi
\left( \mathbf{p},\pm \right) \ ,  \tag{5}
\end{equation}
and, having taken the finite normalization volume, the anticonmutation
relation for the corresponding annihilation and creation operators is

\begin{equation}
\left\{ c\left( \mathbf{p},\pm \right) ,c^{\dagger }\left( \mathbf{p%
{\acute{}}%
},\pm \right) \right\} =\delta _{\mathbf{p\,},\,\mathbf{p\,%
{\acute{}}%
}^{\,}}\ .  \tag{6}
\end{equation}

The phase between helicity states has been chosen to satisfy

\begin{equation}
\pm \sigma ^{2}\chi ^{\star }\left( \mathbf{p},\mp \right) =\chi \left(
\mathbf{p},\pm \right) \ .  \tag{7}
\end{equation}
Convenient basis for a fermion moving in the $x\ y$ plane are

\begin{equation}
\chi \left( \mathbf{p},+\right) =\frac{1}{\sqrt{2}}\left(
\begin{array}{c}
1 \\
-1
\end{array}
\right) \qquad ,\qquad \chi \left( \mathbf{p},-\right) =\frac{-i}{\sqrt{2}}%
\left(
\begin{array}{c}
1 \\
1
\end{array}
\right) \ ,  \tag{8}
\end{equation}

\begin{equation}
\chi \left( \mathbf{p\,%
{\acute{}}%
},+\right) =\frac{1}{\sqrt{2}}\left(
\begin{array}{c}
e^{i\theta /2} \\
-e^{-i\theta /2}
\end{array}
\right) \qquad ,\qquad \chi \left( \mathbf{p\,%
{\acute{}}%
},-\right) =\frac{-i}{\sqrt{2}}\left(
\begin{array}{c}
e^{i\theta /2} \\
e^{-i\theta /2}
\end{array}
\right) \ ,  \tag{9}
\end{equation}
respectively for the initial state when it comes from the positive $x$ axis
and the final one where it is scattered with angle $\theta $ .

\section{Neutrino scattering by string in friction and scaling regimes}

From Eq. (2) the interaction of $\nu _{R}$ of mass $M_{o}$ with the string
is given by

\begin{gather}
\mathcal{L}_{int}=\frac{e}{2}\psi ^{\dagger }\sigma ^{\mu }\mathcal{A}_{\mu
}\psi -\frac{M_{o}}{2}\psi ^{\dagger }\sigma ^{2}\psi ^{\star }\left(
1-e^{i\varphi }f\,\left( \rho \right) \right)  \notag \\
-\frac{M_{o}}{2}\psi ^{T}\sigma ^{2}\psi \left( 1-e^{-i\varphi }f\,\left(
\rho \right) \right) \ ,  \tag{10}
\end{gather}
where, because of the terms of interaction with the Higgs field, the
perturbative method will be applicable for $p>M_{o}$ .

In the string rest frame, the cross-section due to an interaction time $%
\mathcal{T}$ will be

\begin{equation}
\sigma =\sum_{\text{final\thinspace \ states}}\frac{\mathcal{V}}{\mathcal{T}}%
\frac{p_{o}}{p}\left| S_{f\ i}\right| ^{2}\ ,  \tag{11}
\end{equation}
where the sum over final states includes momenta and helicities.

For elastic scattering, the S-matrix \ elements will receive three
contributions

\begin{equation}
S_{f\ i}=S_{f\ i}^{(1)}+S_{f\ i}^{(2)}+S_{f\ i}^{(3)}\ ,  \tag{12}
\end{equation}
where the first corresponds to the interaction with the gauge field and the
others to those with Higgs field. For the perturbative evaluation of Eq.(12)
we will approximate the behaviour of these bosonic classical field Eq.(3) of
the string whose core radius is $R\mathcal{\,\,}\backsim \eta ^{-1}$ as
\begin{gather}
f(r)=a(r)=0\ ,\qquad r<R  \tag{13} \\
f(r)=a(r)=1\ ,\qquad r>R\ .  \tag{14}
\end{gather}
The gauge field contribution to the scattering from positive to positive
helicity fermion in first order perturbation considering Eqs.(4,8,9) gives

\begin{gather}
S_{+,+}^{(1)}=i\int d^{4}x\ \left\langle \mathbf{p\,%
{\acute{}}%
},+\left| \frac{e}{2}\psi ^{\dagger }\sigma ^{\mu }\mathcal{A}_{\mu }\psi
\right| \mathbf{p},+\right\rangle  \notag \\
=\frac{ieL}{2\mathcal{V}}\sqrt{\frac{\left( p_{o}%
{\acute{}}%
+p\,%
{\acute{}}%
\,\,\right) \left( p_{o}+p\right) }{2p_{o}%
{\acute{}}%
\,\,2p_{o}}}2\pi \delta \left( p_{o}%
{\acute{}}%
-p_{o}\right) \ A_{+,+}\ \ ,  \tag{15}
\end{gather}
for a length $L$ of the string and where
\begin{equation}
A_{+,+}=\int drd\varphi \ e^{-i\mathbf{Q.r}}\ \frac{ia(r)}{2e}\left(
e^{i\theta /2+i\varphi }-e^{-i\theta /2-i\varphi }\right) \ .  \tag{16}
\end{equation}
Using the expansion in plane waves
\begin{equation}
e^{-i\mathbf{Q.r}}=\sum_{l=-\infty }^{\infty }\left( -i\right) ^{l}\
J_{l}\left( Q\,r\right) \,e^{-il\left( \varphi -\beta \right) }\ ,  \tag{17}
\end{equation}
where $\beta $ is the angle between the momentum transfer $\mathbf{Q=p%
{\acute{}}%
-p}$ with $x$ axis, only the $l=1$ contribution remains through the
integration over $\varphi $ in Eq.(16) . The subsequent integration over $r$
considering the approximation of Eqs.(13,14) and taking an infrared cutoff $%
\xi $ \ gives

\begin{equation}
A_{+,+}=\frac{2\pi i}{e}\int_{R}^{\xi }dr\ J_{1}\left( Q\,r\right) \ .
\tag{18}
\end{equation}

The first stage following the string formation corresponds to the friction
regime at the beginning of which the correlation length is $\xi =\frac{1}{%
\lambda \eta }$ . The total cross-section per unit length can be then
computed numerically giving the results of Fig.1 . For large enough $\lambda
^{-1}>8$ , and GUT scale $\eta =10^{15}\ GeV$ , the results can be fitted by

\begin{equation}
\frac{d\sigma _{AB}}{dL}=1.86\,\xi \ +\frac{5.52}{p}\ ,  \tag{19}
\end{equation}
showing that the ordinary Aharonov-Bohm behaviour of the second term becomes
to be overrun by the cutoff contribution of the first one. It may be seen
that in the friction regime, though the cone in the forward direction which
gives rise to the cutoff is very relevant, the non forward contribution
cannot be neglected.

For a final state with negative helicity, it turns out that $S_{-,+}^{\left(
1\right) }=0$ \ which could be expected since the Aharonov-Bohm scattering
conserves the helicity \cite{7}. On the other hand the cross-section for
negative to negative helicity is $\backsim (\frac{M_{o}^{2}}{2p^{2}})^{2}$
smaller than Eq.(19) due to the fact that $\nu _{R}$ is essentially of
positive helicity.

Regarding the contribution of the string Higgs field to the scattering of a
positive to positive helicity $\nu _{R}$ , using Eqs.(8,9) ,

\begin{gather}
S_{+,+}^{(2)}=-i\frac{M_{o}}{2}\int d^{4}x\ \left\langle \mathbf{p\,%
{\acute{}}%
},+\left| \psi ^{\dagger }\sigma ^{2}\psi ^{\star }\left( 1-e^{i\varphi
}f\,\,\right) \right| \mathbf{p},+\right\rangle   \notag \\
=\frac{iM_{o}L}{2\mathcal{V}}\sqrt{\frac{\left( p_{o}%
{\acute{}}%
+p\,%
{\acute{}}%
\,\,\right) \left( p_{o}-p\right) }{2p_{o}%
{\acute{}}%
\ 2p_{o}}}\,2\pi \delta \left( p_{o}%
{\acute{}}%
-p_{o}\right) \cos \left( \theta /2\right) \,\frac{\pi }{2p^{2}}\ \digamma
_{+,+}\ ,  \tag{20}
\end{gather}
where
\begin{equation}
\digamma _{+,+}=\int d\varphi \,dr\,r\,e^{-i\mathbf{Q.r}}\left(
1-e^{i\varphi }\ f\,\right) .  \tag{21}
\end{equation}

\begin{figure}[t]
\caption[Figura de Helder Chávez] {\label{fig1}\small Contribution
of the gauge field of the string to the $+\rightarrow
+\,\;$helicity scattering cross section in the friction regime for
different values of $\lambda .$ The lines represent the fit using
Eq.(19).}
\end{figure}

The use of the expansion Eq.(17) and the integration over $\varphi $ leaves
\begin{equation}
\digamma _{+,+}=2\pi \int dr\,r\,\left[ J_{0}\left( Q\,r\right) +i\,f\,\
e^{i\beta }\,J_{1}\left( Q\,r\right) \right] =\frac{\pi }{2p^{2}}\Xi \ ,
\tag{22}
\end{equation}
where, together with the aproximation Eqs.(13,14) and cutoff $\xi $ ,

\begin{gather}
\Xi =\int_{o}^{2p\xi }dz\,z\,J_{0}\left( z\,\cos \left( \theta /2\right)
\right) +i\,e^{i\beta }\int_{2pR}^{2p\xi }dz\,z\,J_{1}\left( z\,\cos \left(
\theta /2\right) \right) \   \notag \\
=\Xi _{o}+i\,\Xi _{1}e^{i\beta }\ .  \tag{23}
\end{gather}
With change from positive to negative helicity the calculation is analogous
giving

\begin{equation}
S_{-,+}^{(2)}=i\frac{M_{o}L}{2\mathcal{V}}\sqrt{\frac{\left( p_{o}%
{\acute{}}%
-p\,%
{\acute{}}%
\,\,\right) \left( p_{o}-p\right) }{2p_{o}%
{\acute{}}%
\ 2p_{o}}}\ 2\pi \delta \left( p_{o}%
{\acute{}}%
-p_{o}\right) \sin \left( \theta /2\right) \,\frac{\pi }{2p^{2}}\ \Xi \ .
\tag{24}
\end{equation}

\begin{figure}[t]
\caption[Figura de Helder Chávez] {\label{fig2}\small Contribution
of the Higgs field of the string to the $\;+$ $\rightarrow -$ \
helicity scattering cross-section in the friction regime for
different values of $\lambda .$ The lines represent the
approximation of Eq.(30).}
\end{figure}

For the matrix element of $S^{\left( 3\right) }$ without change of helicity,
again using Eqs.(8,9),

\begin{gather}
S_{+,+}^{(3)}=-i\frac{M_{o}}{2}\int d^{4}x\ \left\langle \mathbf{p\,%
{\acute{}}%
},+\left| \psi ^{T}\sigma ^{2}\psi ^{\star }\left( 1-e^{-i\varphi }\
f\,\,\right) \right| \mathbf{p},+\right\rangle  \notag \\
=\frac{iM_{o}L}{2\mathcal{V}}\sqrt{\frac{\left( p_{o}%
{\acute{}}%
-p\,%
{\acute{}}%
\,\,\right) \left( p_{o}+p\right) }{2p_{o}%
{\acute{}}%
\ 2p_{o}}}\,2\pi \delta \left( p_{o}%
{\acute{}}%
-p_{o}\right) \cos \left( \theta /2\right) \,\mathcal{G}_{+,+}\ \ ,  \tag{25}
\end{gather}
where now
\begin{equation}
\mathcal{G}_{+,+}=\int d\varphi \,dr\,r\,e^{-i\mathbf{Q.r}}\left(
1-e^{-i\varphi }\ f\,\right) .  \tag{26}
\end{equation}
With the same steps as above one gets
\begin{gather}
\mathcal{G}_{+,+}=2\pi \int \,dr\,r\,\left[ J_{0}\left( Q\,r\right) +i\,f\
e^{-i\beta }J_{1}\left( Q\,r\right) \right]  \notag \\
=\frac{\pi }{2p^{2}}\left( \Xi _{o}+ie^{-i\beta }\,\Xi _{1}\right) \ .
\tag{27}
\end{gather}
Analogously, for the change of helicity

\begin{gather}
S_{-,+}^{(3)}=\frac{iM_{o}L}{2\mathcal{V}}\sqrt{\frac{\left( p_{o}%
{\acute{}}%
+p\,%
{\acute{}}%
\,\,\right) \left( p_{o}+p\right) }{2p_{o}%
{\acute{}}%
\ 2p_{o}}}2\pi \delta \left( p_{o}%
{\acute{}}%
-p_{o}\right) \sin \left( \theta /2\right) \,  \notag \\
\times \ \frac{\pi }{2p^{2}}\left( \Xi _{o}+ie^{-i\beta }\,\Xi _{1}\right) \
.  \tag{28}
\end{gather}

Using these results
\begin{equation}
\left| S_{+,+}^{(2)}+S_{+,+}^{(3)}\right| ^{2}\backsim \left( \frac{M_{o}}{p}%
\right) ^{2}\left| S_{-,+}^{(2)}+S_{-,+}^{(3)}\right| ^{2},  \tag{29}
\end{equation}
indicating the fact that violation of helicity is favoured by the Majorana
coupling.

The numerical computation of this dominant cross-section is shown in Fig.2
and can be fitted for $\lambda ^{-1}\gtrsim 8$\ by the approximate behaviour
for $p\xi \gg 1$

\begin{equation}
\frac{d\sigma _{\mathcal{H}}}{dL}=1.04\,\xi \left( \frac{M_{o}}{p}\right)
^{2}[1+0.48\ln \left( p\xi \right) ]\ ,  \tag{30}
\end{equation}
where the additional logarithmic dependence on the cutoff is a consequence
of the phase in its interaction with the fermion as seen in Eq.(10).

For a later time after their formation, strings may reach \cite{8} the
scaling regime, due to a correlation length $\xi \backsim t$ , when the
universe cooled below $T_{sc}\simeq \frac{T_{GUT}^{2}}{M_{pl}}\backsim
10^{11}GeV.$ \ This occurred for the time $t\backsim 10^{-28}s$ , being the
expansion of the universe scale due to radiation $a(t)\varpropto t^{1/2}$ .
Therefore for momentum $p>M_{o}$ , $\xi p\gg 1$ \ with $\xi \gg R$ so that
now the cutoff contribution dominates clearly over the ordinary
Aharonov-Bohm term giving as approximation for the cross-section caused by
the gauge field

\begin{equation}
\frac{d\sigma _{AB}}{dL}\simeq 2\left( 1+\frac{M_{o}^{2}}{4p^{2}}\right)
^{2}\xi \ ,  \tag{31}
\end{equation}
and obviously the contribution given by the string Higgs field is even
better approximated by Eq.(30) in the scaling regime. Since the correlation
length is much larger in the scaling regime than in the friction one, the
above cross-sections are correspondingly larger in the former case.

\section{Capture of fermions by strings with emission of bosons}

This process has analogy with the capture of an electron by a nucleus with
the emission of a photon, where the description is given in terms of the
interaction of the electron with the quantized radiation field in addition
to the Coulomb attraction.

Therefore we add the quantum fluctuations to the classical configurations of
the string Higgs and gauge fields as

\begin{equation}
\Phi =\Phi ^{cl}+\widehat{\Phi }\qquad ,\qquad \mathcal{A}_{\mu }=\mathcal{A}%
_{\mu }^{cl}+\widehat{\mathcal{A}}_{\mu }\ .  \tag{32}
\end{equation}
Thus we will have as interaction with the additional quantum boson fields

\begin{equation}
\mathcal{L}^{\text{quan}}=-\frac{ie}{2}\psi ^{\dagger }\sigma ^{\mu }\,%
\widehat{\mathcal{A}}_{\mu }\psi -\frac{ig}{2}\left( \psi ^{\dagger }%
\widehat{\Phi }\psi ^{C}-\psi ^{C^{\ \dagger }}\widehat{\Phi }^{\dagger
}\psi \right) \ ,  \tag{33}
\end{equation}
for which the conditions for the validity of the perturbation treatment are $%
e\lesssim 1$ and $g\lesssim 1.$

Now for the fermion field we must consider the free solutions and the
zero-mode states which can be formed with the background of the classical
string configuration, i.e.

\begin{equation}
\psi =\widehat{\psi }_{free}+\widehat{\psi }_{zm}\ ,  \tag{34}
\end{equation}
where $\widehat{\psi }_{free}$ is given by Eq. (4) as before whereas the
zero mode term will be

\begin{equation}
\widehat{\psi }_{zm}=\sum_{p_{z}\,>\,0}\left[ c_{o}\left( p_{z},+\right) \,%
\mathcal{U}_{o}\left( p_{z},+\right) \,e^{-i\omega t}+c_{o}^{\dagger }\left(
p_{z},+\right) \,\mathcal{U}_{o}^{\,\star }\left( p_{z},+\right)
\,e^{\,i\omega t}\right] \ ,  \tag{35}
\end{equation}
with $\omega =p_{z}$ , describing massless particles which move along the
positive z-axis through the anticommuting operators $c_{o}\,.$ The zero-mode
wave funtion is given \cite{9} by

\begin{equation}
\mathcal{U}_{o}\left( p_{z},+\right) =\frac{\widetilde{M}}{\sqrt{2\pi L}}%
\left(
\begin{array}{c}
1 \\
0
\end{array}
\right) \exp \left( -\int_{0}^{\rho }\left[ \frac{M_{o}}{M_{\mathcal{H}}}%
f\,\left( \rho \,%
{\acute{}}%
\,\,\,\right) +\frac{a\left( \rho \,%
{\acute{}}%
\,\,\,\right) }{2\rho \,%
{\acute{}}%
}\right] \,d\rho \,%
{\acute{}}%
\,\right) \exp \left( ip_{z}\,z\right) \ ,  \tag{36}
\end{equation}
where $\widetilde{M}^{-1}$ gives its effective radius, $\rho
{\acute{}}%
=M_{\mathcal{H}}\,r$ \ and the normalization is $\int d^{3}x\,\left|
\mathcal{U}_{o}\left( p_{z},+\right) \right| ^{2}=1/2$ . It is clear that
this requires a more detailed description of the classical fields inside the
string than that given by Eqs.(13,14), i.e. \cite{10}
\begin{equation}
f\left( \rho \,%
{\acute{}}%
\right) =f_{o}\,\rho \,%
{\acute{}}%
\qquad ,\qquad a\left( \rho
{\acute{}}%
\right) =a_{o}\,\rho
{\acute{}}%
\,^{2}\qquad ,\ \rho \,%
{\acute{}}%
<1\ ,  \tag{37}
\end{equation}
$a_{o}$ and $f_{o}$ being constants that, from the normalization condition,
give $\widetilde{M}=M_{\mathcal{H}}\sqrt{\frac{M_{o}}{M_{\mathcal{H}}}f_{o}+%
\frac{a_{o}}{2}}$ . The boson quantum fields outside the string are massive
and are described in terms of operators with usual commutators for the
complex Higgs and real gauge ones as

\begin{gather}
\widehat{\Phi }(x)=\sum_{\overrightarrow{k}}\frac{1}{\sqrt{2k_{o}\mathcal{V}}%
}\left( a_{k}e^{-ik.x}+b_{k}^{\dagger }e^{ik.x}\right) \ ,  \tag{38} \\
\widehat{\mathcal{A}}_{\mu }(x)=\sum_{\lambda }\sum_{\mathbf{p}}\frac{1}{%
\sqrt{2p_{o}\mathcal{V}}}\left( \varepsilon _{\mu }(p,\lambda )\,a(p,\lambda
)e^{-ip.x}+\varepsilon _{\mu }^{\star }(p,\lambda )\,a^{\dagger }(p,\lambda
)e^{ip.x}\right) \ ,  \tag{39}
\end{gather}
the polarization vectors satisfying

\begin{equation}
\varepsilon _{\mu }^{\star }(p,\lambda )\varepsilon ^{\mu }(p,\lambda \,%
{\acute{}}%
\,\,)=-\delta _{\lambda \lambda \,%
{\acute{}}%
}\qquad ,\qquad \sum_{\lambda }\varepsilon _{\mu }(p,\lambda )\varepsilon
_{\nu }^{\star }(p,\lambda )=-\,g_{\mu \nu }+\frac{p_{\mu }p_{\nu }}{M_{%
\mathcal{A}}^{2}}\ .  \tag{40}
\end{equation}

The capture of a neutrino with emission of a superheavy Higgs particle $\nu
_{R}\rightarrow \nu _{zm}+\Phi $ is produced by the second term of Eq. (33)
giving at first order of perturbation theory the probability amplitude

\begin{equation}
\mathcal{S}_{zm\,\Phi ,\nu _{R}}=\frac{g}{4}\frac{\widetilde{M}}{\mathcal{V}%
\sqrt{\pi q_{o}L}}\sqrt{\frac{p_{o}+p}{2p_{o}}}2\pi \delta \left( p_{o}-q_{o}%
{\acute{}}%
-q_{o}\right) 2\pi \delta \left( q_{z}%
{\acute{}}%
+q_{z}\right) \,\Gamma \left( Q\right) \ ,  \tag{41}
\end{equation}
where

\begin{equation}
\Gamma \left( Q\right) =\frac{1}{\sqrt{2}}\int d^{2}x\,e^{i\mathbf{x}_{T}%
\mathbf{.Q}}\exp \left( -\int_{0}^{\rho }\left[ \frac{M_{o}}{M_{\mathcal{H}}}%
f\,\left( \rho \,%
{\acute{}}%
\,\,\right) +\frac{a(\rho \,%
{\acute{}}%
\,\,\,)\ }{2\rho \,%
{\acute{}}%
}\right] \,d\rho \,%
{\acute{}}%
\right) \ .  \tag{42}
\end{equation}
This integral in the transverse plane of the string, with the momentum
transferred to it $\mathbf{Q=p-q}_{T}%
{\acute{}}%
$ can be calculated approximately \cite{9} expanding the plane wave in
Bessel functions to give
\begin{equation}
\int d\varphi \ e^{i\mathbf{x}_{T}\mathbf{.Q}}=2\pi \,J_{0}\left(
Q\,r\right) \ ,  \tag{43}
\end{equation}
which together with the form of the classical fields inside the string
Eq.(37), being neglegible their contribution outside it, allows to obtain

\begin{equation}
\Gamma \left( Q\right) \simeq \sqrt{2}\pi \frac{1}{\widetilde{M}^{2}}\exp
\left( -\frac{Q^{2}}{2\widetilde{M}^{2}}\right) \ .  \tag{44}
\end{equation}
The numerical evaluation of the capture total cross-section is shown in
Fig.3 taking all the masses of the same order.

Regarding the capture with the emission of one gauge vector boson $\nu
_{R}\rightarrow \nu _{zm}+\mathcal{A}$ , the first term of Eq. (33) gives
the probability amplitude

\begin{gather}
\mathcal{S}_{zm\,\mathcal{A},\nu _{R}}=\frac{ie}{2\mathcal{V}}\frac{%
\widetilde{M}}{\sqrt{\pi L}}\sqrt{\frac{p_{o}+p}{2p_{o}2k_{o}}}2\pi \delta
\left( p_{o}-k_{o}-q_{o}\right) 2\pi \delta \left( q_{z}+k_{z}\right)  \notag
\\
\times \,\,\chi _{o}^{\dagger }\sigma ^{\mu }\varepsilon _{\mu }^{\star
}(k,\lambda )\,\chi \left( \mathbf{p},+\right) \Gamma \left( Q\right) \ ,
\tag{45}
\end{gather}
where now $Q=\left| \mathbf{p}-\mathbf{k}_{T}\right| $ , $\chi _{o}=\left(
\begin{array}{c}
1 \\
0
\end{array}
\right) ,$ $\chi \left( \mathbf{p},+\right) =\frac{1}{\sqrt{2}}\left(
\begin{array}{c}
1 \\
-1
\end{array}
\right) $ and $\Gamma \left( Q\right) $ is that of Eq.(42) .

The numerical results for the total cross-section with the same
approximation of equal masses and taking \cite{11} $\alpha _{GUT}^{-1}=\frac{%
4\pi }{e^{2}}=26$ are also presented in Fig. 3 .

One sees that both cross-sections show a resonant behaviour for values of
the momentum of $\nu _{R}$ of the order of its mass .

\begin{figure}[t]
\caption[Figura de Helder Chávez] {\label{fig2}\small Comparison
of the cross sections of capture of $\nu _{R}$ to form zero modes
with emission either of a Higgs or a vector massive boson .}
\end{figure}

\section{Cosmological implications and conclusions}

We have analysed the possibility that fermions that acquired mass in the GUT
epoch of the universe evolution could have been captured by cosmic strings
formed by the breaking of an Abelian subgroup at this scale.

In the case that the GUT symmetry that contained this subgroup corresponded
to $SO(10)$ , the fermion to be considered is the $\nu _{R}$ . These
neutrinos captured by the string would produce a superconducting current,
even though they are neutral, in the sense that inside it they travel at the
velocity of light.

This current stabilizes closed strings which otherwise would contract and
disappear. The superconducting microscopic loops, vortons, detach from the
string dynamics and behave as a quasi-stable dark matter that might have
survived till the present condensed in the halo of our galaxy. Since vortons
have a small probability of quantum decay by tunneling of their
constituents, they might be the origin of the ultra-high energy cosmic rays
(UHECR) which have been observed without identification of their
astrophysical sources \cite{12} .

To see which is the flux of UHECR produced by vortons it is necessary to
estimate their density whose evolution with the temperature of universe $T$
starting from their formation at $T_{f}$ will be
\begin{equation}
n_{v}(T)=n(T_{f})\left( \frac{T}{T_{f}}\right) ^{3},  \tag{46}
\end{equation}
being considered \cite{13} that $n(T_{f})\backsim \left( \xi (T_{f})\right)
^{-3}.$

During the friction regime an estimation \cite{14} is
\begin{equation}
\xi \,^{fr}(T)\backsimeq (m_{PL})^{1/2}\frac{T_{GUT}}{T\,^{5/2}}\ ,  \tag{47}
\end{equation}
where $m_{PL}$ is the Plank mass, so that for $T_{f}=T_{GUT}$ from Eq.(46) $%
n_{v}^{\,fr}(T)\backsimeq 10^{-6}\,T^{3}$ , whereas for the formation at the
end of this period $T_{f}\backsimeq 10^{11}GeV$ \ $n_{v}^{\,fr}(T)\backsimeq
10^{-24}\,T^{3}.$

Regarding the number of fermionic carriers in the loop \cite{13}
\begin{equation}
N\backsimeq \xi (T_{f})T_{GUT}\,,  \tag{48}
\end{equation}
in the friction regime $N\,^{fr}\backsim 100$ if $T_{f}=T_{GUT}$ , and $%
N\,^{fr}\backsim 10^{12}$ at the end of it when $T_{f}\backsimeq 10^{11}GeV.$

Looking now at the formation in the scaling regime valid for $T\lesssim
10^{11}GeV$ where
\begin{equation}
\xi ^{sc}\backsimeq H^{-1}\backsimeq \frac{m_{PL}}{T^{2}}\ ,  \tag{49}
\end{equation}
being $H$ the Hubble parameter, $n_{v}^{sc}(T)\lesssim 10^{-24}T^{3}$ and $%
N^{sc}>10^{12},$ both in agreement with the limit of the friction epoch.

As a consequence the number of fermions in vortons per unit volume $N\
n_{v}(T)$ if incorporated at the beginning of friction regime is $%
10^{-4}T^{3},$ whereas if incorporated at the beginning of the scaling one
will be $10^{-12}T^{3}.$ Therefore the ratio of these incorporated fermions
per unit comoving volume is $10^{8}$ , which is equal to the inverse ratio
of times of formation. A similar analysis for the vorton formation during
the scaling regime indicates that the density of absorbed fermions per unit
comoving volume goes like $\;t_{f}^{-1/2}.$

It is obvious that the above ratio of fermions equals the one of lengths of
original closed strings which at formation will be $\xi \left( T_{f}\right)
\times \xi ^{-3}\left( T_{f}\right) .$ Therefore, being the probability of
capture per unit length independent on time, the formation of vortons should
be equally probable in friction and scaling regimes. However if one includes
the motion of the original strings, which is different in both regimes, the
formation of vortons during the scaling period is less likely \cite{15} .

Even without this last consideration, from the above estimation of vortons
density it is clear that their relevance for UHECR will be much more
important if they were formed at the beginning of the friction epoch.

The considered example of $SO(10)$ is the simplest one since $\nu _{R}$ is
the only non-ordinary fermion and acquires mass at the GUT scale. In order
to have exotic fermions electrically charged and which have zero modes
giving way to superconducting currents in the common sense of the word, we
should consider the unification of interactions under a larger group as $%
E_{6}$ . But in this case the addition of $11$ fermions, apart from $\nu
_{R} $ , would make the analysis of the problem considerably harder.

Regarding the scattering of fermions by straight and long strings one may
note that to the traditional Aharonov-Bohm effect due to the gauge
potential, as in the solenoid case, also the interaction with Higgs field
which generates the fermion mass must be added. From our calculation, one
sees that this effect in the total scattering cross-section increases with
the separation among strings faster than that due to the gauge field by a
logarithmic factor which is related to the winding phase present also at
large distances.

For this kind of strings their density length will be $\backsim 1/\xi ^{2}$
and, subtracting the universe expansion, the corresponding one per unit
comoving volume in the scaling regime will go like $\;1/t$ . Since the
cross-section per unit string length for scattering of neutrinos increases
at least as $t$ , the effect due to the targets in unit comoving volume will
be roughly constant.

To make the process of generation of superconducting currents more
realistic, one should take into account the propagation of neutrinos in the
plasma outside the string, the influence of the motion of the latter and the
fluctuations of the field equivalent to the electric one which could produce
jumps of the fermions from negative to positive energy inside the core.

\noindent

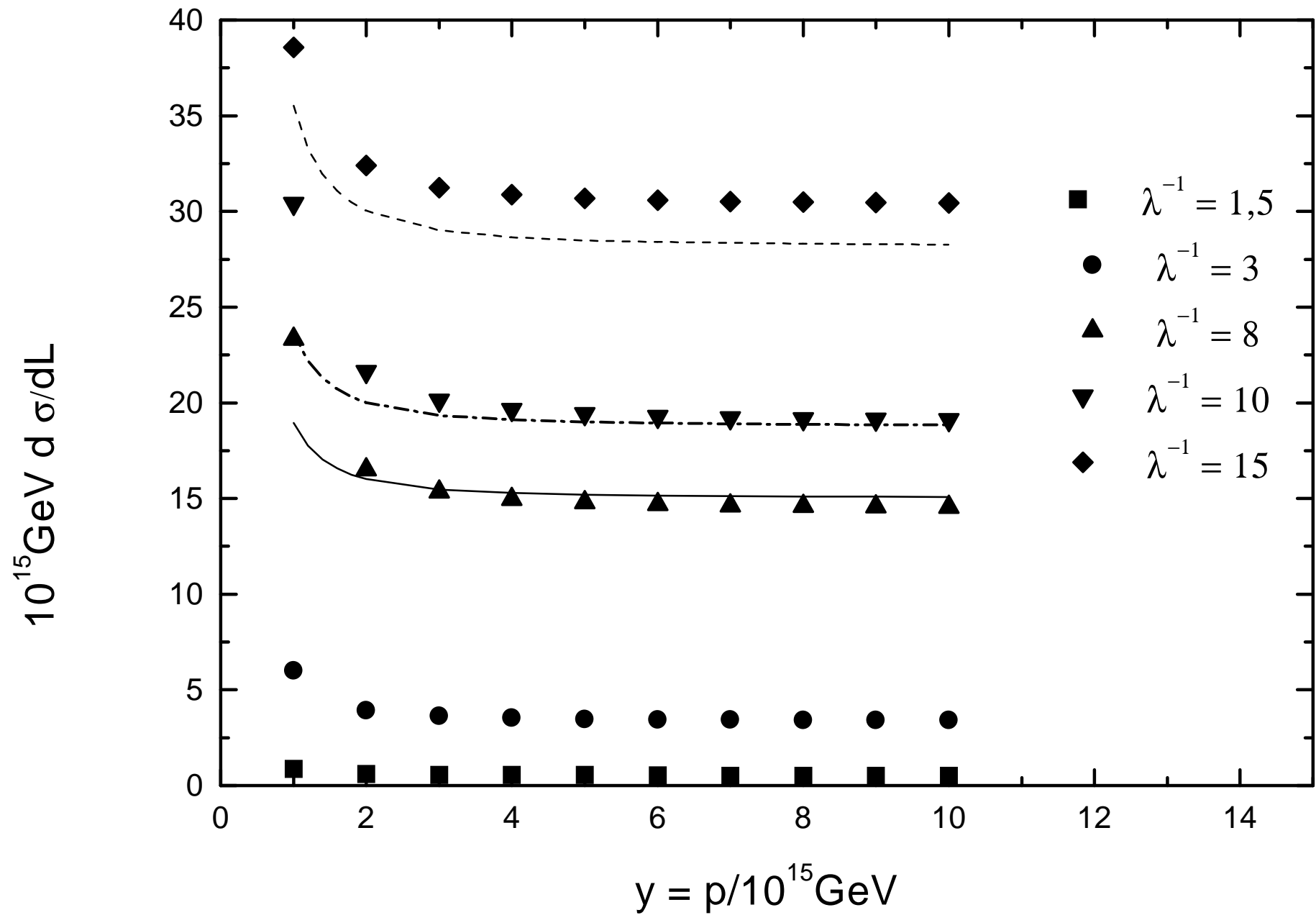
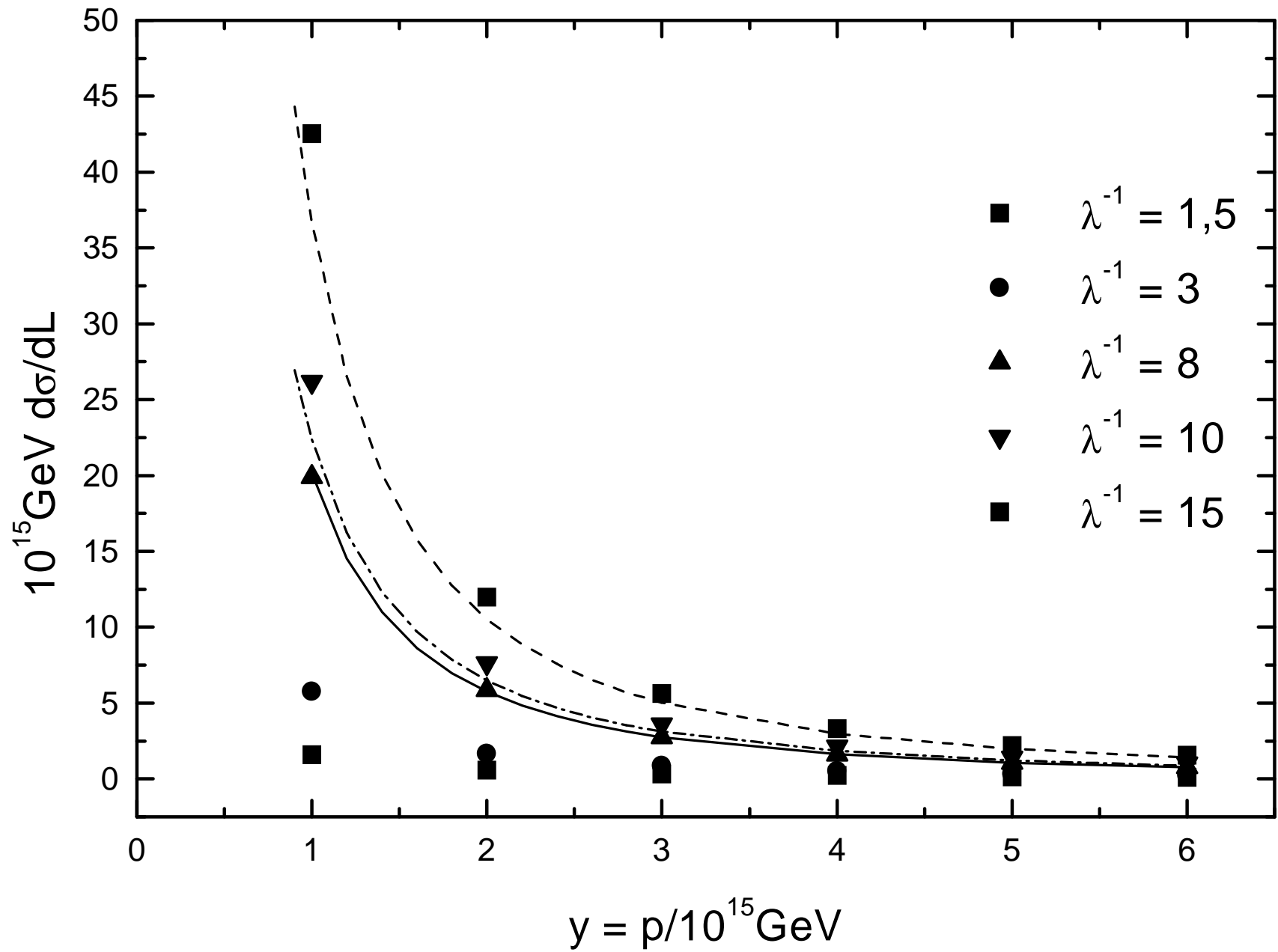
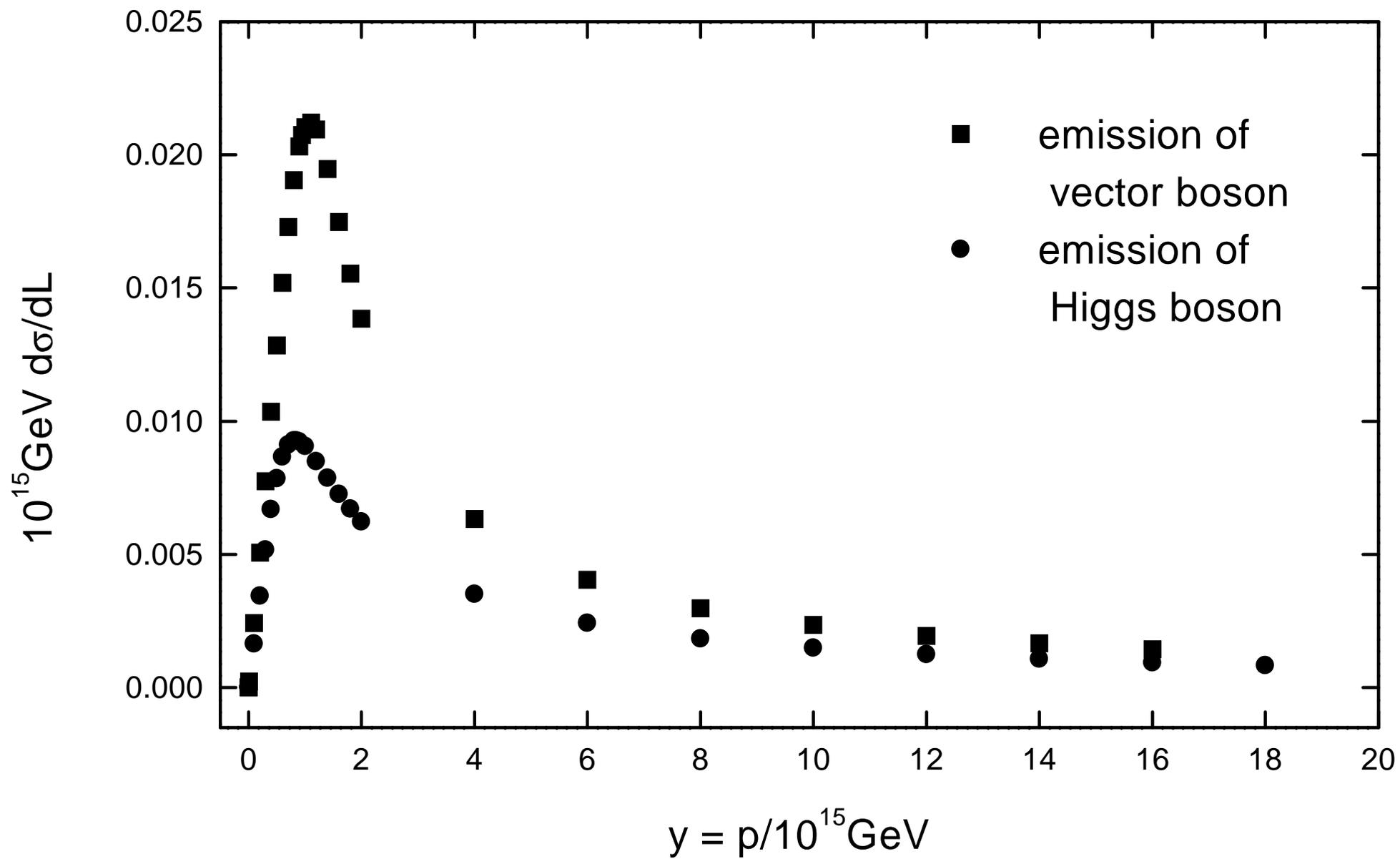
\end{document}